\documentclass[%
print,
 amsmath,amssymb,
 aps,
]{revtex4}

\usepackage[sc]{mathpazo}
\usepackage{graphicx}
\usepackage{dcolumn}
\usepackage{bm}
\usepackage[landscape,papersize={297.1mm,210mm},left=1.9cm,right=1.6cm,top=2.7cm,bottom=2.8cm]{geometry}
\usepackage[                   
            pdfstartview=FitH,
            colorlinks, 
            pdfborder=001,   
            linkcolor=blue,
            anchorcolor=blue,
            citecolor=blue,
            urlcolor=blue
            ]{hyperref}

\begin{document}
\title{Necessary and sufficient product criteria for quantum states via the rank of realignment matrix of density matrix}

\author{Xianfei Qi}
\author{Ting Gao}
\email{gaoting@hebtu.edu.cn}
\affiliation {College of Mathematics and Information Science, Hebei
Normal University, Shijiazhuang 050024, China}
\author{Fengli Yan}
\email{flyan@hebtu.edu.cn}
\affiliation {College of Physics Science and Information Engineering, Hebei
Normal University, Shijiazhuang 050024, China}

\begin{abstract}
We present a necessary and sufficient product criterion for bipartite quantum states based on the rank of realignment matrix of density matrix. Then, this approach is generalized to multipartite systems. We first introduce the concept of semiproduct in a similar manner to the semiseparable and prove that semiproduct is equivalent to fully product. Therefore, a quantum state is bipartite product with respect to all possible partitions implies fully product which is different from the case of separability. For pure states, it can easily be seen that several necessary and sufficient separability criteria for multipartite systems are derived as a special case of our results. Several specific examples illustrate that our criteria are convenient and operational.
\end{abstract}


\maketitle

\section{Introduction}
Quantum entanglement has been regarded as an important physical resource in the theory of quantum information over the past three decades \cite{RMP81.865}. It plays a vital role in quantum information processing and has widely been applied to many fields such as quantum cryptography \cite{PRL67.661}, quantum teleportation \cite{PRL70.1895,EPL84.50001}, quantum dense coding \cite{PRL69.2881} and quantum metrology \cite{Nature2010}. So, it has theoretical and practical value to study the entanglement of quantum states.

Quantum states can be divided into separable states and entangled states. Although it is known to be NP-hard for the problem of separability of states \cite{JCSC2004}, a considerable number of separability criteria have been proposed. For bipartite systems, there are many famous separability criteria \cite{RMP81.865}. For multipartite systems, this problem become more complicated owing to the complicated structure of multipartite entangled states. For example, there are only two cases in bipartite systems, separable or entangled, while, in multipartite quantum systems, we have to face the problem of $k$-partite entanglement or $k$-nonseparability for a given partition and an unfixed partition. Despite the difficulty, many efficient separability criteria for multipartite systems \cite{QIC2008,QIC2010,PRA82.062113,EPL104.20007,PRA91.042313,SR5.13138,PRA93.042310,EPJD61.765} and computable measures \cite{PRA83.062325,PRA86.062323,PRL112.180501} quantifying multipartite entanglement have been presented.

Quantum states can also be divided into product states and non-product states. In the following, we discuss the relation between product states and separable states. For bipartite pure states, the definition of product states coincides with the definition of separable states, that is, bipartite pure separable states are product states and vice versa. While, a bipartite mixed state is called separable if it can be expressed as a convex combination of product states. Otherwise, it is called entangled. Note that, unlike the special case of pure states, the set of separable states is in general strictly larger than the set of product states \cite{1612.02437}. Physically, a product state is an uncorrelated state. For non-product states, there are two different kinds of correlation, classical correlation and quantum correlation. Separable states are classically correlated. The reason is that a separable state can be prepared just by local operations and classical communication (LOCC) \cite{PR474.1}. On the contrary, entangled states are quantum correlated. An enormous amount of researches have been devoted to quantum entanglement \cite{RMP81.865} and other quantum correlation beyond entanglement \cite{RMP84.1655,JPA49.473001,1703.10542}.

In this paper, we investigate the product criteria for quantum states. First, we prove that the rank of realignment matrix of density matrix equals one is the necessary and sufficient condition of this density matrix being a product state for bipartite systems. Then, this result is generalized to the case of multipartite systems. As a special case, we can see that these criteria are necessary and sufficient for determining $k$-separability or $k$-nonseparability of a multipartite pure state.

This paper is organized as follows. In Sec.~\uppercase\expandafter{\romannumeral 2} we introduce basic concepts and notations that will be used in the subsequent sections. In Sec.~\uppercase\expandafter{\romannumeral 3}, we provide our central result, that is, the close relation between the rank of realignment matrix of density matrix and the property of product for bipartite states. In Sec.~\uppercase\expandafter{\romannumeral 4}, we discuss the property of product for multipartite quantum states as a generalization of the case of bipartite quantum systems. Sec.~\uppercase\expandafter{\romannumeral 5} is conclusion.

\section{Preliminaries}
Before we state the main results, we introduce some necessary knowledge about the matrix realignment. For a matrix $X=(x_{ij})\in \mathbb{C}^{m\times n}$, the vector \text{vec}($X$) is defined as
$$\text{vec}(X)=[x_{11},\ldots,x_{m1},x_{12},\ldots,x_{m2},\ldots,x_{1n},\ldots,x_{mn}]^{\text{T}},$$
where T denotes the transpose. Let $Z$ be an $m\times m$ block matrix with each block $Z_{i,j}\in \mathbb{C}^{n\times n}$, the realigned matrix $Z^{R}$ is defined by
$$Z^{R}=[\text{vec}(Z_{11}),\ldots,\text{vec}(Z_{m1}),\text{vec}(Z_{12}),\ldots,\text{vec}(Z_{m2}),\ldots,\text{vec}(Z_{1m}),\ldots,\text{vec}(Z_{mm})]^{\text{T}}.$$

\emph{Lemma 1} \cite{Matrix,Ph.D} For a matrix $Z\in \mathbb{C}^{mn\times mn}$, if $Z^{R}\in \mathbb{C}^{m^{2}\times n^{2}}$ of rank $r$ has the singular value decomposition $Z^{R}=U\sum V^{\dag}=\sum\limits_{i=1}^{r}\sigma_{i}u_{i}v_{i}^{\dag}$, where $\sum=\left(
                                                                                                           \begin{array}{cc}
                                                                                                             \sum_{r} & 0 \\
                                                                                                             0 & 0 \\
                                                                                                           \end{array}
                                                                                                         \right)\in \mathbb{C}^{m^{2}\times n^{2}},
\sum_{r}=\text{diag}\{\sigma_{1},\sigma_{2},\ldots,\sigma_{r}\}\in \mathbb{C}^{r\times r}$ with the nonzero singular values $\sigma_{i}~(i=1,2,\ldots,r)$ of $Z^R$ in nonincreasing order,  then, $Z$ can be expressed as
\begin{equation}
\begin{aligned}
Z=\sum\limits_{i=1}^{r}(X_{i}\otimes Y_{i}),\label{0}
\end{aligned}
\end{equation}
with vec($X_{i}$)=$\sqrt{\sigma_{i}}u_{i}$ and vec($Y_{i}$)=$\sqrt{\sigma_{i}}v_{i}^{*}$, where $U=[u_{1},u_{2},\cdots, u_{m^{2}}]\in \mathbb{C}^{m^{2}\times m^{2}}$ and $V=[v_{1},v_{2},\cdots, v_{n^{2}}]\in \mathbb{C}^{n^{2}\times n^{2}}$ are unitary.

\emph{Lemma 2} \cite{Ph.D}
A matrix $Z$ can be expressed as the tensor product of two matrices $X$ and $Y$, that is,
\begin{equation}
\begin{aligned}
Z=X\otimes Y~~~ \text{if and only if}~~~ Z^{R}=\text{vec}(X)\text{vec}(Y)^{\text{T}}.\label{1}
\end{aligned}
\end{equation}

In this paper, we consider a multiparticle quantum system with state space $\mathcal{H}=\mathcal{H}_{1} \otimes \mathcal{H}_{2} \otimes \ldots \otimes \mathcal{H}_{N}$, where $\mathcal{H}_{i}~(i=1,2,\ldots,N)$ denote $d_{i}$-dimensional Hilbert spaces. A quantum state is called fully product state if can be written as tensor product of states of each subsystems,
\begin{equation}
\begin{aligned}
\rho=\rho_{1}\otimes\rho_{2}\otimes\cdots\otimes\rho_{N}.
\end{aligned}
\end{equation}

Next, we introduce the definition of $k$-product state. To this end, we first introduce the notion of $k$-partition \cite{EPL104.20007,PRA86.062323}.
A $k$-partition $A_{1}\mid A_{2}\mid \cdots \mid A_{k}$ (of $\{1,2,\ldots,N\})$ means that the set $\{A_{1},A_{2},\ldots,A_{k}\}$ is a collection of pairwise disjoint subsets of $\{1,2,\ldots,N\}$ with the union of all subsets $A_{i}$ is $\{1,2,\ldots,N\}$ (pairwise disjoint union $\bigcup_{i=1}^{k}A_{i}=\{1,2,\ldots,N\}$). An $N$-partite state is called $k$-product if there is a $k$-partition $A_{1}\mid A_{2}\mid \cdots \mid A_{k}=j_{1}^{1}\cdots j_{n_{1}}^{1}\mid j_{1}^{2}\cdots j_{n_{2}}^{2}\mid \cdots \mid j_{1}^{k}\cdots j_{n_{k}}^{k}$ such that
\begin{equation}
\begin{aligned}
\rho=\rho_{A_{1}}\otimes\rho_{A_{2}}\otimes\cdots\otimes\rho_{A_{k}},
\end{aligned}
\end{equation}
where $\rho_{A_{i}}$ is the state of subsystems $A_{i}$ and disjoint union $\bigcup_{t=1}^{k}A_{t}=\bigcup_{t=1}^{k}\{j_{1}^{t},j_{2}^{t}\cdots j_{n_{t}}^{t}\}=\{1,2,\cdots,N\}$. Obviously, the definition of $k$-product states is equivalent to the definition of $k$-separable states for pure states \cite{EPL104.20007,PRA86.062323}.

\section{Product criterion for bipartite states via the rank of realignment matrix of density matrix}
In this section, we are ready to present the central result of the article. Theorem 1 provides a necessary and sufficient product criterion for bipartite quantum systems by using the rank of realignment matrix of density matrix. It demonstrates that a bipartite mixed state is product state if and only if the rank of realignment matrix of density matrix equals one. A simple example clearly shows the difference between product states and separable states for mixed states.

\emph{Lemma 3}~~If an $mn\times mn$ positive matrix $P$ can be written as the tensor product of an $m\times m$ matrix $\widetilde{M}$ and an $n\times n$ matrix $\widetilde{N}$, i.e., $P=\widetilde{M}\otimes \widetilde{N}$, then there exist $m\times m$ positive matrix $M$ and $n\times n$ positive matrix $N$ such that $P=M\otimes N$.

\emph{Proof}~~Let $P$ be an $mn\times mn$ positive matrix and $P=\widetilde{M}\otimes \widetilde{N}$, then for $\forall$ $|\psi\rangle\in C^{m\times 1}$, $|\phi\rangle\in C^{n\times 1}$, we have
\begin{equation}
\begin{aligned}
(|\psi\rangle\otimes |\phi\rangle, \widetilde{M}\otimes \widetilde{N}|\psi\rangle\otimes |\phi\rangle)=\langle\psi|\widetilde{M}|\psi\rangle\langle\phi|\widetilde{N}|\phi\rangle\geqslant 0.\label{2}
\end{aligned}
\end{equation}
Therefore, both $\widetilde{M}$ and $\widetilde{N}$ must be positive matrix or negative matrix simultaneously. Otherwise, there exist $|\psi_{1}\rangle, |\psi_{2}\rangle\in C^{m\times 1}$, $|\phi_{1}\rangle, |\phi_{2}\rangle\in C^{n\times 1}$  such that
$\langle\psi_{1}|\widetilde{M}|\psi_{1}\rangle> 0, \langle\phi_{1}|\widetilde{N}|\phi_{1}\rangle> 0$, while $\langle\psi_{2}|\widetilde{M}|\psi_{2}\rangle< 0, \langle\phi_{2}|\widetilde{N}|\phi_{2}\rangle< 0$, which implies that $(|\psi_{1}\rangle\otimes |\phi_{2}\rangle, \widetilde{M}\otimes \widetilde{N}|\psi_{1}\rangle\otimes |\phi_{2}\rangle)< 0$. This contradicts the fact that $P$ is a positive matrix. If $\widetilde{M}$ and $\widetilde{N}$ are negative matrices, we can define $M=-\widetilde{M}$ and $N=-\widetilde{N}$. We have thus proved the lemma.

\emph{Theorem 1}~~An $mn\times mn$ density matrix $\rho$ can be expressed as the tensor product of an $m\times m$ density matrix $\rho_{1}$ and an $n\times n$ density matrix $\rho_{2}$, i.e., $\rho=\rho_{1}\otimes \rho_{2}$, if and only if $r(\rho^{R})=1$.

\emph{Proof}~~Let $\rho$ is an $mn\times mn$ density matrix and there exist an $m\times m$ density matrix $\rho_{1}$ and an $n\times n$ density matrix $\rho_{2}$ such that $\rho=\rho_{1}\otimes \rho_{2}$. Note that, a matrix is rank one  iff it can be written as product of a column vector and a row vector. Combining this property with (\ref{1}), we have $r(\rho^{R})=1$.

On the other hand, if $r(\rho^{R})=1$, there are matrices $\widetilde{\rho_{1}}$ and $\widetilde{\rho_{2}}$ such that $\rho=\widetilde{\rho_{1}}\otimes \widetilde{\rho_{2}}$. Because density matrix is a positive matrix, according to Lemma 3, there exist positive matrices $\rho{'}_{1}$ and $\rho{'}_{2}$ such that $\rho=\rho{'}_{1}\otimes \rho{'}_{2}$. Besides,
$\textrm{Tr}(\rho)=\textrm{Tr}(\rho{'}_{1}\otimes \rho{'}_{2})=\textrm{Tr}(\rho{'}_{1})\cdot \textrm{Tr}(\rho{'}_{2})=1$. Let $\rho_{1}=\rho{'}_{1}/\textrm{Tr}(\rho{'}_{1})$, $\rho_{2}=\rho{'}_{2}/\textrm{Tr}(\rho{'}_{2})$, then the traces of $\rho_{1}$ and $\rho_{2}$ both are one. That is to say, $\rho_{1}$ and $\rho_{2}$ are density matrices and $\rho=\rho_{1}\otimes \rho_{2}$.

\emph{Example 1.} Consider the family of two qubit states,
$$\rho=p\frac{|00\rangle\langle 00|+|11\rangle\langle 11|}{2}+\frac{1-p}{4}I_{4},$$
where $I_{4}$ is the $4\times 4$ identity matrix. Obviously, $\rho$ is separable because it is a mixture of a separable state and white noise.
The realignment matrix of density matrix is
$$\left(
  \begin{array}{cccc}
    \frac{1+p}{4} & 0 & 0 & \frac{1-p}{4} \\
    0 & 0 & 0 & 0 \\
    0 & 0 & 0 & 0 \\
    \frac{1-p}{4} & 0 & 0 & \frac{1+p}{4} \\
  \end{array}
\right).
$$
It is easy to see that when $p\in(0,1]$, $\rho$ is non-product state. This simple example clearly illustrates that the set of separable states is in general strictly larger than the set of product states.

\section{Product criteria for multipartite states via the rank of realignment matrix of density matrix}

In the following, a generalization to multipartite system is presented. First we define the concept of semiproduct in a similar manner to the semiseparable \cite{RMP81.865}. An $N$-partite quantum state is called semiproduct if it is product under all $1$ versus $N-1$ partitions $i|\overline{i} : =i|1,\ldots,i-1,i+1,\ldots,N, 1\leqslant i\leqslant N$.

\emph{Theorem 2}~~Let $\rho$ be a $d_{1}d_{2}\cdots d_{N}\times d_{1}d_{2}\cdots d_{N}$ density matrix, then the following statements are  equivalent:

(1) $\rho$ is fully product;

(2) $\rho$ is semiproduct;

(3) $\rho$ is product under all possible 2-partitions.

\emph{Proof} ~ Since it is trivial that (1) $\Rightarrow$ (3) and (3) $\Rightarrow$ (2), the proof will  therefore be complete when we have shown that
(2) $\Rightarrow$ (1). Suppose that $\rho$ is semiproduct, we prove that $\rho$ is fully product by induction. For $N=3$,  $\rho=\rho_{1}\otimes\rho_{23}=\rho_{2}\otimes\rho_{13}$. By tracing out subsystem 1, we get $\text{Tr}_{1}(\rho)=\rho_{23}=\rho_{2}\otimes\text{Tr}_{1}(\rho_{13})$, i.e., $\rho_{23}=\rho_{2}\otimes\rho_{3}$ with $\rho_{3}=\text{Tr}_{1}(\rho_{13})$. Here $\rho_{3}$ is a density matrix from Lemma 3. Assume that the conclusion is also true for $k$, then for $k+1$, we have $\rho=\rho_{1}\otimes\rho_{\overline{1}}=\rho_{2}\otimes\rho_{\overline{2}}=\cdots=\rho_{k+1}\otimes\rho_{\overline{k+1}}$, where $\rho_{i}$ is a  $d_{i}\times d_{i}$ density matrix and $\rho_{\overline{i}}$ is a $d_{1}d_{2}\cdots d_{i-1}d_{i+1}\cdots d_{k+1}\times d_{1}d_{2}\cdots d_{i-1}d_{i+1}\cdots d_{k+1}$ density matrix, $i=1,2,\ldots,k+1$. Tracing out the last subsystem, we find $\text{Tr}_{k+1}(\rho)=\rho_{1}\otimes\text{Tr}_{k+1}(\rho_{\overline{1}})=\cdots=\rho_{k}\otimes \text{Tr}_{k+1}(\rho_{\overline{k}})=\rho_{\overline{k+1}}$. Based on the assumption, we know $\rho_{\overline{k+1}}$ is a fully product state of $k$ subsystems. Therefore, $\rho$ is a fully product state of $k+1$ subsystems.

In the end, it is easy to see that the next Theorem 3 follows immediately from Theorem 2.

\emph{Theorem 3}~~Let $\rho$ be a $d_{1}d_{2}\cdots d_{N}\times d_{1}d_{2}\cdots d_{N}$ density matrix, $\rho$ is a $k$-product state iff there exists a $k$-partition $A_{1}\mid A_{2}\mid \cdots \mid A_{k}$ such that $\rho$ is a semiproduct state under this $k$-partition  iff there exists a $k$-partition $A_{1}\mid A_{2}\mid \cdots \mid A_{k}$ such that $\rho$ is product by taking into account all possible bipartite partitions under this $k$-partition.

For multipartite pure states, Theorem 3 can also be seen as providing a necessary and sufficient separability criterion for determining $k$-separability or $k$-nonseparability of a multipartite pure state.

\emph{Example 2.} Let us consider the three-qubit state, $|\psi\rangle=\frac{|0_{1}0_{2}0_{3}\rangle+|1_{1}1_{2}0_{3}\rangle}{\sqrt{2}}$. After calculation, the realignment matrix of density matrix $\rho=|\psi\rangle\langle \psi|$ under 2-partition $1|23$ is
$$\left(
    \begin{array}{cccccccccccccccc}
      \frac{1}{2} & 0 & 0 & 0 & 0 & 0 & 0 & 0 & 0 & 0 & 0 & 0 & 0 & 0 & 0 & 0 \\
      0 & 0 & \frac{1}{2} & 0 & 0 & 0 & 0 & 0 & 0 & 0 & 0 & 0 & 0 & 0 & 0 & 0 \\
      0 & 0 & 0 & 0 & 0 & 0 & 0 & 0 & \frac{1}{2} & 0 & 0 & 0 & 0 & 0 & 0 & 0 \\
      0 & 0 & 0 & 0 & 0 & 0 & 0 & 0 & 0 & 0 & \frac{1}{2} & 0 & 0 & 0 & 0 & 0 \\
    \end{array}
  \right).
$$

The realignment matrix of density matrix $\rho=|\psi\rangle\langle \psi|$ under partition $2|13$ is the same as the case of partition $1|23$. Then, according to Theorem 1, $|\psi\rangle$ is not a separable state with respect to the 2-partitions $1|23$ and $2|13$. Thus, $|\psi\rangle$ is not a fully separable state owing to Theorem 2.

The realignment matrix of density matrix $\rho=|\psi\rangle\langle \psi|$ under bipartition $3|12$ is
$$\left(
    \begin{array}{cccccccccccccccc}
      \frac{1}{2} & 0 & 0 & \frac{1}{2} & 0 & 0 & 0 & 0 & 0 & 0 & 0 & 0 & \frac{1}{2} & 0 & 0 & \frac{1}{2} \\
      0 & 0 & 0 & 0 & 0 & 0 & 0 & 0 & 0 & 0 & 0 & 0 & 0 & 0 & 0 & 0 \\
      0 & 0 & 0 & 0 & 0 & 0 & 0 & 0 & 0 & 0 & 0 & 0 & 0 & 0 & 0 & 0 \\
      0 & 0 & 0 & 0 & 0 & 0 & 0 & 0 & 0 & 0 & 0 & 0 & 0 & 0 & 0 & 0 \\
    \end{array}
  \right),
$$
where $|\psi\rangle$ is expressed as $|\psi\rangle=\frac{|0_{3}0_{1}0_{2}\rangle+|0_{3}1_{1}1_{2}\rangle}{\sqrt{2}}$.

According to Theorems 1 and 2, $|\psi\rangle$ is neither a genuinely tripartite entangled state nor a fully separable state, $|\psi\rangle$ is a bipartite separable state under bipartition $3|12$.

\emph{Example 3.} Let us consider the three-qubit W state, $|\psi\rangle=\frac{|0_{1}0_{2}1_{3}\rangle+|0_{1}1_{2}0_{3}\rangle+|1_{1}0_{2}0_{3}\rangle}{\sqrt{3}}$. After calculation, the realignment matrix of density matrix $\rho=|\psi\rangle\langle \psi|$ under bipartition $1|23$ is
$$\left(
    \begin{array}{cccccccccccccccc}
      0 & 0 & 0 & 0 & 0 & \frac{1}{3} & \frac{1}{3} & 0 & 0 & \frac{1}{3} & \frac{1}{3} & 0 & 0 & 0 & 0 & 0 \\
      0 & 0 & 0 & 0 & \frac{1}{3} & 0 & 0 & 0 & \frac{1}{3} & 0 & 0 & 0 & 0 & 0 & 0 & 0 \\
      0 & \frac{1}{3} & \frac{1}{3} & 0 & 0 & 0 & 0 & 0 & 0 & 0 & 0 & 0 & 0 & 0 & 0 & 0 \\
      \frac{1}{3} & 0 & 0 & 0 & 0 & 0 & 0 & 0 & 0 & 0 & 0 & 0 & 0 & 0 & 0 & 0 \\
    \end{array}
  \right)
$$
The realignment matrixes of density matrix $\rho=|\psi\rangle\langle \psi|$ under partition $2|13$ and partition $3|12$ are identical to above matrix because of the highly symmetry that W state possesses. It is concluded that W state is a genuinely tripartite entangled state because it is not separable with respect to all possible 2-partitions by Theorem 1.

\section{Conclusion}
In summary, we reveal the close link between the rank of realignment matrix of density matrix and the property of product for quantum system. We prove that a bipartite quantum state is a product state if and only if the rank of realignment matrix of density matrix equals one. Then, it is naturally generalized to multipartite systems by viewing the total system as the tensor product of different partitions of subsystems. Especially, we prove that semiproduct is equivalent to fully product, which deeply reflects difference between product and separability. As a special case, our results turn into the necessary and sufficient separability criteria for pure states. Several specific examples illustrate that our criteria are convenient and operational. Our result may be beneficial to the study of quantum entanglement.

\acknowledgments
This work was supported by the National Natural Science Foundation of China under Grant Nos: 11371005, 11475054, and Hebei Natural Science Foundation of China under Grant No: A2016205145.

\end{document}